# MASSIVE CLOSE BINARIES

Dany Vanbeveren
*Astrophysical Institute, Vrije Universiteit Brussel, Pleinlaan 2, 1050 Brussels, Belgium*
dvbevere@vub.ac.be

**Abstract**

The evolution of massive stars in general, massive close binaries in particular depend on processes where, despite many efforts, the physics are still uncertain. Here we discuss the effects of stellar wind as function of metallicity during different evolutionary phases and the effects of rotation and we highlight the importance for population (number) synthesis. Models are proposed for the X-ray binaries with a black hole component. We then present an overall scheme for a massive star PNS code. This code, in combination with an appropriate set of single star and close binary evolutionary computations predicts the O-type star and Wolf-Rayet star population, the population of double compact star binaries and the supernova rates, in regions where star formation is continuous in time.

## 1. INTRODUCTION

Population number synthesis (PNS) of massive stars relies on the evolution of massive stars and, therefore, uncertainties in stellar evolution imply uncertainties in PNS. The mass transfer and the accretion process during Roche lobe overflow (RLOF) in a binary, the merger process and common envelope evolution have been discussed frequently in the past by many authors (for extended reviews see e.g. van den Heuvel, 1993, Vanbeveren et al., 1998a, b). In the present paper we will focus on the effects of stellar wind mass loss during the various phases of stellar evolution and the effect of the metallicity Z. The second part of the paper summarises the Brussels overall massive star PNS with a realistic fraction of binaries. We compare theoretical prediction with observations of O-type stars, Wolf-Rayet (WR) stars, supernova (SN) rates and binaries with two relativistic components [neutron star (NS) or black hole (BH)], in regions where star formation is continuous in time. In the last part of the paper, we will discuss overall evolutionary effects of rotation and the consequences for PNS. As usual, we will classify binaries as Case A, Case $B_r$, Case $B_c$ and Case C depending on the onset of RLOF.

## 2. STELLAR WIND FORMALISMS

We distinct four stellar wind (SW) phases: the core hydrogen burning (the OB-type phase) prior to the Luminous Blue Variable (LBV) phase, the LBV phase, the Red Supergiant (RSG) phase, the Wolf-Rayet (WR) phase. To study the effect on evolution one relies on observed SW mass loss rates and how these rates vary as function of stellar parameters (luminosity, temperature, chemical abundance's, rotation). In the following, $\dot{M}$ is in $M_\odot/yr$ and the luminosity L is in $L_\odot$

**The OB-type and LBV phase.** SW mass loss formalisms of OB-type stars have been reviewed by Kudritzki and Puls (2000) (see also P. Crowther, in these proceedings). Note that all formalisms published since 1988 give very similar results when they are implemented into a stellar evolutionary code.

LBVs are hot, unstable and very luminous OB-type supergiants. Those brighter than $M_{bol}$ = -9.5 (corresponding to stars with initial mass ≥ 40 $M_\odot$) are particularly interesting as far as evolution is concerned. They are losing mass by a more or less steady stellar wind ($\dot{M} = 10^{-7}$-$10^{-4}$ $M_\odot/yr$) and by eruptions where the star may lose mass at a rate of $10^{-3}$-$10^{-2}$ $M_\odot/yr$. The outflow velocities are typically a few 100 km/s. The observed SW-rates are highly uncertain, but due to the fact that no RSGs are observed brighter than $M_{bol}$ = -9.5 an evolutionary working hypothesis could be: *the $\dot{M}$ during the LBV + RSG phase of a star with $M_{bol}$ ≤ -9.5 must be sufficiently large to assure a RSG phase which is short enough to explain the lack of observed RSGs with $M_{bol}$ ≤ -9.5.*

**The RSG phase.** Using the formalism proposed by Jura (1987), Reid et al. (1990) determined the SW mass loss rate for 16 luminous RSGs in the Large Magellanic Cloud. The following equation relates these $\dot{M}$ values to the luminosity

$$Log(-\dot{M}) = 0.8 LogL - C \qquad (1)$$

with C = 8.7. However, RSG mass loss rates are subject of possible large uncertainties and C = 9 ($\dot{M}$ is a factor 2 smaller than predicted with C = 8.7) can not be excluded.

**The WR phase.** There is increasing evidence that the outflowing atmospheres of WR stars are inhomogeneous and consist of clumps (Crowther, the present proceedings). SW mass loss rates of WR stars determined by methods that include the effects of clumps are a factor 2-4 times smaller than those obtained with smooth wind models.

Table 1 lists a number of galactic WR stars where the $\dot{M}$ and the luminosity is determined by detailed spectroscopic analysis.

Notice that the spectroscopic $\dot{M}$-value of the WR component of the binary V444 Cyg corresponds to independent determinations which are based on the binary period variation (Khaliulin et al., 1984) and on polarimetry (St.-Louis et al., 1988; Nugis et al., 1998), whereas $\gamma^2$Vel is a nearby WR binary and its distance (based on Hipparcos measurements) is very well known, thus also its luminosity. Figure 1 illustrates a possible $\dot{M}$-L relation. We have given large statistical weight at the observations of V444 Cyg and of γ²Vel. WR 44 is a peak which seems to be unrealistic from statistical point of view. Either the luminosity is too high or the mass loss rate is too small. Omitting WR 44, the following formula links $\dot{M}$ and L

$$Log(-\dot{M}) = LogL - 10 \qquad (2)$$

Remark that including WR 44 gives a relation which predicts even smaller $\dot{M}$-values for Log L ≥ 5 compared to Equation 2

*Table 1*        Luminosity and mass loss rates of galactic WR stars. [1] = De Marco et al., 2000, [2] = Hillier and Miller, 1999, [3] = Morris et al., 2000, [4] = Hamann and Koesterke, 1998, [5] = Nugis et al., 1998}

| WR star | Log L | Log(-$\dot{M}$) | ref |
|---|---|---|---|
| γ²Vel | 5 | -5 | [1] |
| HD 165763 | 5.3 | -4.8 | [2] |
| WR 147 | 5.65 | -4.59 | [3] |
| HD 50896 | 5.45 | -4.42 | [4] |
| V444 Cyg | 5 | -5.04 | [5] |
| WR 44 | 5.55 | -5.1 | [4] |
| WR 123 | 5.7 | -4.14 | [4] |

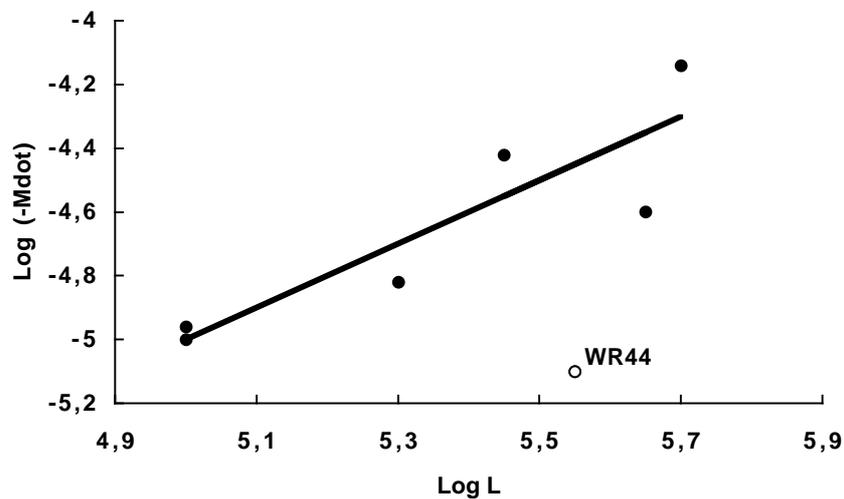

*Figure 1*        The Log $\dot{M}$-Log L diagram of galactic WR stars.

**The effect of metallicity Z on $\dot{M}$** SW rates resulting from spectral analysis are uncertain by at least a factor 2. It is therefore difficult to deduce any kind of metallicity dependence from observations. Theory predicts that for a radiation driven SW, $\dot{M} \div Z^\zeta$ with $0.5 \leq \zeta \leq 1$. We expect that OB-type mass loss rates obey this rule.

We do not know how the SW rates of LBVs and RSGs depend on Z. We like to remind that the $\dot{M}$-L relation (Equation 1) holds for RSGs in the LMC. If the $\dot{M}$ of RSGs depends on Z as well, the effect on the evolution of galactic stars could be even larger.

Hamann and Koesterke (2000) investigated 18 WN stars in the LMC. Figure 2 illustrates the Log $\dot{M}$-Log L diagram. There seems to be no obvious relation however, as an exercise, we used a linear relation similar to (Equation 2) and shifted the line until the sum of the distances between observed values and predicted ones was smallest (a linear regression where the slope of the line is fixed). This resulted in a shift of -0.2 which, accounting for the Z value of the Solar neighbourhood and of the LMC corresponds to $\zeta = 0.5$. To be more conclusive, many more observations are needed where detailed analysis may give reliable stellar parameters, but from this exercise it looks as if *WR mass loss rates are Z-dependent*.

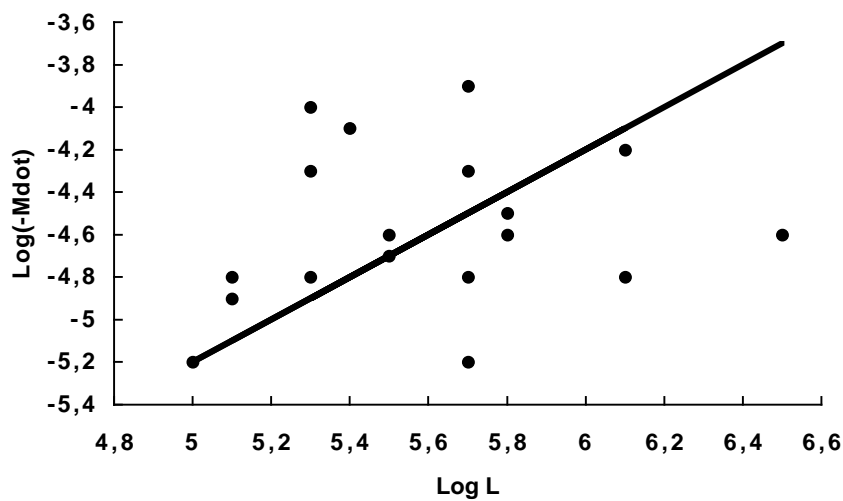

*Figure 2*     The Log $\dot{M}$-Log L diagram for WR stars in the LMC.

## 3. MASSIVE STAR EVOLUTION

When the most recent SW mass loss formalisms of OB-type stars are implemented into a stellar evolutionary code, one concludes *that stellar wind mass loss during core hydrogen burning (prior to the LBV phase) is not important for the evolution of galactic stars with initial mass smaller than 40 $M_\odot$*.

A star with initial mass larger than 40 $M_\odot$ loses most of its hydrogen rich layers by OB+LBV-type stellar wind. When it is a close binary component and when the binary period P is large enough so that RLOF does not start before the onset of the LBV phase (roughly P ≥ 10 days), mass loss by RLOF may be largely suppressed (or may not occur at all), i.e. *due to OB-type + LBV-type stellar wind, the importance of RLOF in a binary with orbital period ≥ 10 days and with primary mass larger than 40 $M_\odot$ may be significantly reduced; RLOF may even be avoided*.

Of particular importance is the evolution of the binary period due to stellar wind. The specific angular momentum ($\dot{J}/\dot{M}$) of a spherically symmetric wind can be expressed as

$$\frac{\dot{J}}{\dot{M}} = \lambda A^2 \Omega \qquad (3)$$

with A the distance between the two binary components and $\Omega = \frac{2\pi}{P}\Omega$. When the wind velocity of the mass loser (with mass $M_1$ and mass loss $\dot{M}_1$) is large compared to the orbital velocity $\Omega A$ of the companion (with mass $M_2$), this specific angular momentum can be approximated by the momentum of matter at the moment it leaves the loser, i.e.

$$\lambda = \left(\frac{M_2}{M_1 + M_2}\right)^2 \qquad (4)$$

resulting in an overall binary period increase.

However, when the wind velocity is comparable to the orbital velocity, the outflowing matter can gain orbital angular momentum by torque and the value of $\lambda$ may be significantly larger than predicted by Equation 4. This is particularly important for the evolution of binaries with very small mass ratio, possible progenitors of the Low Mass X-ray Binaries (LMXBs) with a BH component (see Sect. 4)

### 3.1. THE EFFECT OF RSG MASS LOSS

The influence of RSG mass loss rates (Equation 1) on single star evolution has been described in extenso by Vanbeveren et al., (1998a, b) and Salasnich et al. (1999). One of the important consequences of the larger RSG rates is that *Galactic single stars with initial mass as low as 20(15) $M_\odot$ become hydrogen deficient core helium burning stars = WR stars*.

Hamann and Koesterke (1998, 2000) determined the luminosity and temperature of a large sample of WR stars in the Galaxy and in the LMC. When one considers the WN stars without hydrogen and if at least some of the WR stars with the lowest luminosity are single, their position in the HR-diagram is indirect evidence for the RSG-wind evolution of single stars as discussed in the papers cited above.

When a star with initial mass ≥ 15-20 $M_\odot$ is a binary component and when the period is large enough so that RSG wind starts before the star fills its Roche lobe (Case $B_c$ and Case C binaries), it is clear that RSG wind can significantly reduce the importance of the RLOF mass loss (and the common envelope process). Even more, case C will not happen. Let us remark that an RSG wind is very slow. This means that λ in Equation 3 may be substantially larger than predicted by Equation 4

If RSG mass loss rates are Z-dependent as predicted by the radiatively driven wind theory, it can readily be understood that the minimum mass of single WR star progenitors in small Z-regions will be larger than in the Galaxy.

## 3.2. THE EFFECT OF WR MASS LOSS

Due to LBV and/or RSG stellar wind mass loss or due to RLOF, a massive star may become a hydrogen deficient core helium burning star. The further evolution is governed by a WR-type stellar wind mass loss. The effect on massive star evolution of present day WR-type mass loss rates (Equation 2) has been studied by Vanbeveren et al. (1998 a, b, c). Recently, Wellstein and Langer (1999) discussed the core helium burning (CHeB) evolution of massive stars by adopting a stellar wind mass loss rate relation

$$Log(-\dot{M}) = 1.5 Log L - 12.25 \qquad (5)$$

for hydrogen deficient CHeB stars with logL ≥ 4.45. Compared to Equation 2, two differences are important. Equation 5 assumes a stronger luminosity dependence whereas it predicts $\dot{M}$-values for the stars of table 1 which are still significantly larger than the observed values. We will compare our CHeB results with results calculated with Equation 5.

**The Galactic WR lifetime.** To deduce the WR-lifetime one obviously needs a definition of a WR star that is useful for evolution. Most of the single WR stars have Log L ≥ 5 (Hamann and Koesterke, 1998, 2000) whereas most of the WR components in WR+OB binaries have a mass ≥ 8 $M_\odot$ (Vanbeveren et al., 1998b). Accounting for the theoretical M-L relation of WR stars discussed later, it can be concluded that both conditions are similar. We will use the Log L ≥ 5 criterion. Figure 3 shows the galactic WN and WC lifetimes as a function of initial progenitor mass. We only consider the core helium burning WR stars. The most massive stars may become WNL stars (WN stars with significant hydrogen in their atmosphere) already during core hydrogen burning (see Andre Maeder, present proceedings) and this obviously increases the total WN lifetime compared to the lifetimes shown in Figure 3. Remark that most of the WNE (WN stars without hydrogen) stars originate from stars with initial mass between 20 $M_\odot$ and 40 $M_\odot$ whereas most of the WC stars have M ≥ 40 $M_\odot$ progenitors. The figures illustrate that a PNS study of WR stars will not critically depend on whether evolutionary results are used where the RSG stellar wind mass loss is computed with Equation 1 with C = 8.7 or with C = 9. The different wind formalisms during the WR phase have a larger effect (typically a factor 2).

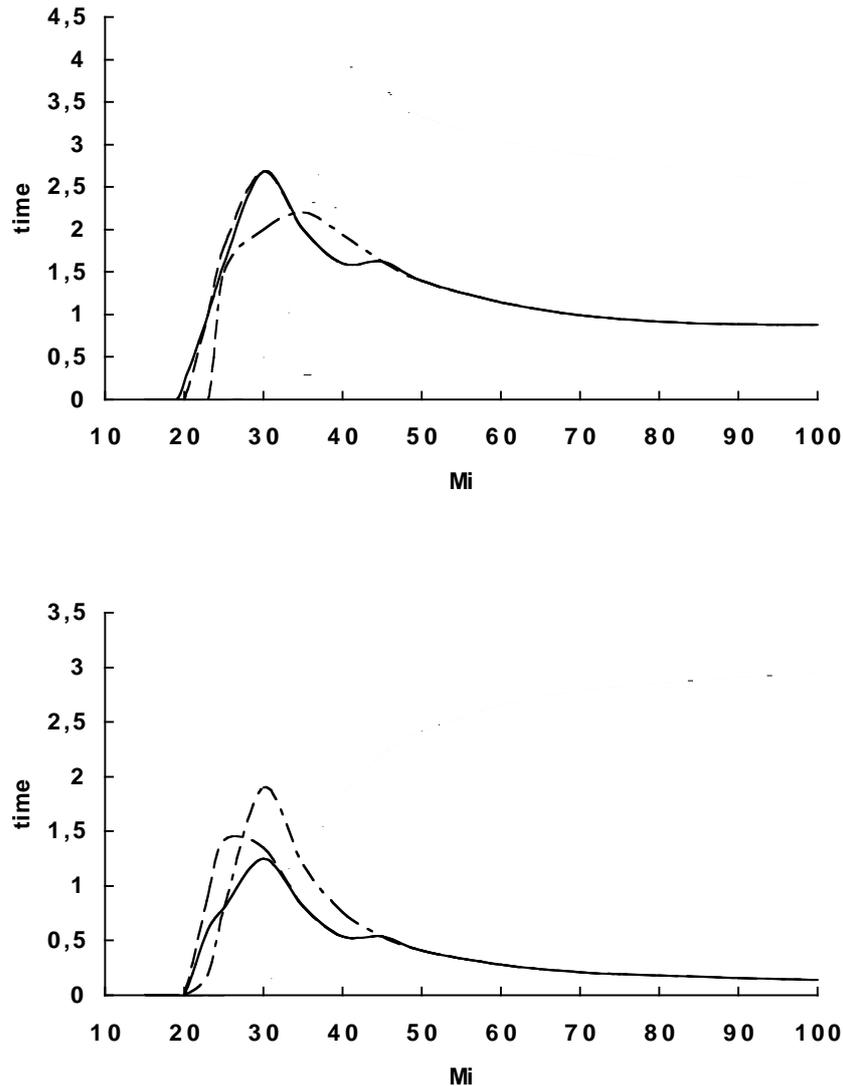

*Figure 3*     The WN (thick lines) and WC (thin lines) lifetime (in $10^5$ yr) as function of initial progenitor mass (in $M_\odot$). Dashed-dotted (respectively dashed lines and full lines) lines apply to case B binaries (respectively single stars where the RSG mass loss is calculated with Equation 1 with C = 8,7 and with C = 9). Above 40 $M_\odot$ all stars lose their hydrogen rich layers very fast due to the LBV-type formalism discussed in Sect. 2. The top-figure (respectively bottom-figure) corresponds to our preferred WR mass loss formalism Equation 2 (respectively Equation 5).

**The M-L relation of WR stars.**   A mass luminosity relation of WR stars has been proposed by Vanbeveren and Packet (1979), Maeder (1983), Langer (1989), Vanbeveren (1991), Schaerer and Maeder (1992). All relations closely match, although different assumptions were made concerning a number of uncertainties during CHeB. This relation can therefore be considered as a very robust one.

**The final mass prior to core collapse.** Figure 4 gives the final mass of the star and of the CO-core prior to core collapse, for single stars and for Case B binary components. Remind that Case B is the more frequent class of interacting binaries. Compared to Case B, it is trivial to understand that Case A components have smaller final masses whereas it is obvious that Case C (and non-interacting) binaries are similar to single stars. Equation 1 decides upon the mass loss during the RSG phase. The top-figure holds for the WR rates predicted by Equation 2 whereas the bottom-figure corresponds to the formalism adopted by Wellstein and Langer (2000). Notice the much larger final masses when Equation 2 holds, i.e. *with our preferred WR mass loss formalism, stars with initial mass between 40 $M_\odot$ and 100 $M_\odot$ end their life with a mass between 10 $M_\odot$ and 20 $M_\odot$ corresponding to CO-cores with a mass between 5 $M_\odot$ and 15 $M_\odot$.*

It must be clear that PNS of binaries with relativistic components, NSs or BHs, depends critically on the adopted WR mass loss formalism.

Figure 4 can be used to link the minimum mass of BH formation of single stars with the one of primaries of case B-type binaries. Suppose that the latter is 40 $M_\odot$. Based on the value of the CO-core mass, the corresponding minimum mass for single stars is ~30 $M_\odot$ (RSG stellar wind according to Equation 1 with C = 8.7) or ~20 $M_\odot$ (Equation 1 with C = 9). Obviously, these mass limits also apply for Case C binaries and for Case $B_c$ ones where the RSG wind phase begins before the onset of the RLOF (Sect. 3.1).

**Case BB versus stellar wind.** Habets (1985, 1986) computed the evolution of helium stars with $2 \leq M/M_\odot \leq 4$ up to neon ignition, assuming that the SW mass loss during CHeB is small in this mass range. He concluded that CHeB stars with $2 \leq M/M_\odot \leq 2.9$ expand significantly during the He shell burning phase, after CHeB. If a star like that is a close binary component, it can be expected that it will overflow the critical Roche lobe for a second time: the process is known as case BB RLOF. However, when we extrapolate the stellar wind mass loss formalism (Equation 2) and we calculate the evolution of CHeB stars with $2 \leq M/M_\odot \leq 2.9$, it follows that case BB is largely suppressed (or does not happen at all). It can readily be checked that the difference between case BB and SW mainly concerns the period evolution of binaries.

## 4. APPLICATIONS

**ν Sgr.** The binary has a period P = 138 days, a hydrogen deficient A-type supergiant with minimum mass = 2.5 $M_\odot$ and a B-type companion with minimum mass = 4 $M_\odot$ (Dudley and Jefferey, 1990). A 2.5 $M_\odot$ helium core corresponds to a primary with initial mass ~12 $M_\odot$ and we propose a 12 $M_\odot$ + 4 $M_\odot$ progenitor binary. The present period indicates that the initial period was very large, corresponding to Case $B_c$ or Case C binaries, and this means that the binary experienced common envelope evolution. When it is assumed that all the energy that is needed to remove the hydrogen rich layers of the 12 $M_\odot$ star comes from the orbit, it is impossible to explain the present binary period. We propose an alternative model: when we use the RSG stellar wind mass loss formalism

(Sect. 2), the star may lose ~5-6 M$_\odot$ by stellar wind before the onset of the common envelope. Orbital energy is then responsible for the removal of the remaining 3-4 M$_\odot$ and for the reduction of the binary period to 138 days. We conclude that the binary v Sgr gives indirect evidence for the importance of RSG stellar wind mass loss in massive stars, possibly down to 10-12 M$_\odot$.

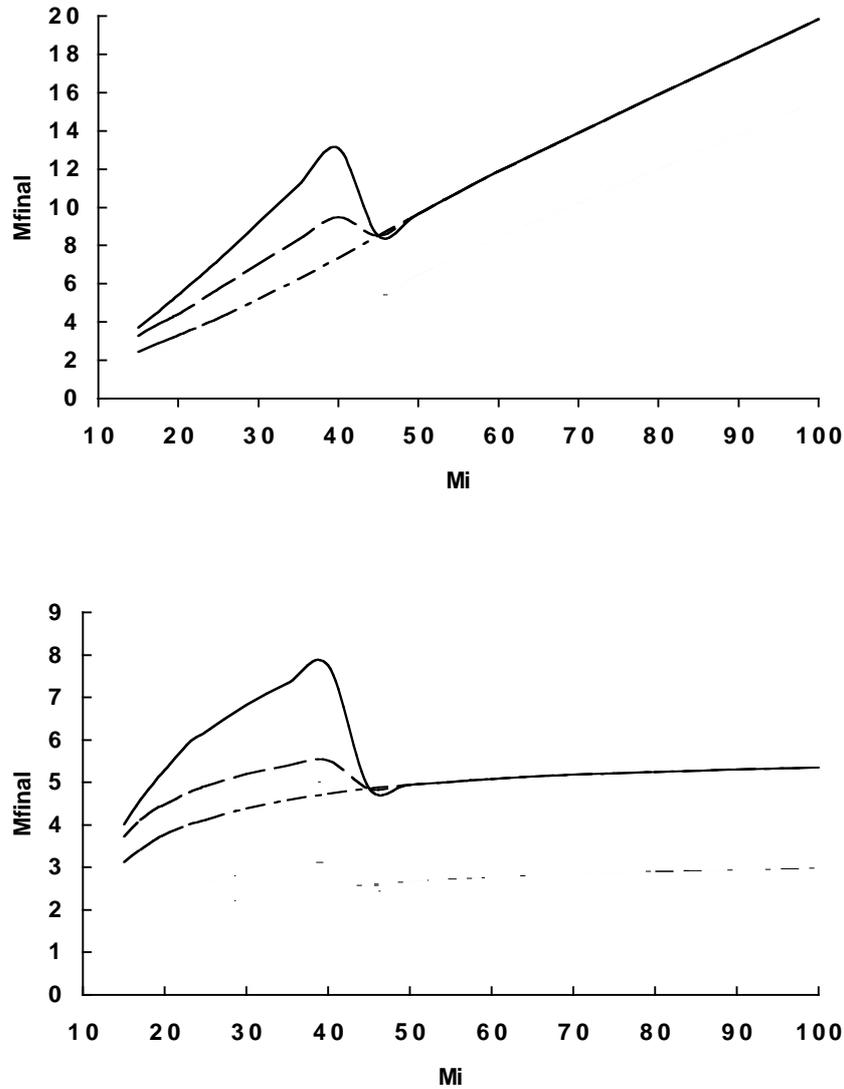

*Figure 4* The final mass (in M$_\odot$) of the star (thick lines) and of the CO core (thin lines) as a function of initial mass (in M$_\odot$) of the star. The line-types have the same meaning as in Figure 3. The top-figure (respectively bottom-figure) corresponds to our preferred WR mass loss formalism Equation 2 (respectively Equation 5)

**Cyg X-1.** The compact star in the high mass X-ray binary Cyg X-1 is most likely a BH with mass ~10 $M_\odot$ (Gies and Bolton, 1986; Herrero et al., 1995). Nelemans et al. (1999) demonstrated that the space velocity of the binary (~50 km/s) can be explained if ~4 $M_\odot$ were ejected during a previous supernova explosion. This means that the final mass of the BH progenitor was at least 14 $M_\odot$. Figure 4 (top-figure) illustrates that the initial mass of the BH progenitor was larger than 60-70 $M_\odot$. With the WR mass loss rates proposed by Wellstein and Langer, it is not possible to explain the BH mass. The initial mass indicates that LBV+RSG stellar wind mass loss has played a fundamental role in the evolution of the progenitor system (Sect. 3 and Sect. 3.1).

**The LMXBs with a BH component.** Bailyn et al. (1998) list a number of LMXBs with a (probable) BH component. The BH masses range between 2 $M_\odot$ and 18 $M_\odot$ (7 $M_\odot$ is the average), the binary period between 0.2 days and 6 days, the mass of the optical component between 0.3 $M_\odot$ and 2.6 $M_\odot$. To explain the space velocity of Nova Sco 1994 (≥ 100 km/s), Nelemans et al. (1999) propose that ~4 $M_\odot$ of the BH progenitor was lost during the supernova explosion. If this value can be considered as some canonical one, the average final mass of the BH progenitors was ~11 $M_\odot$. Figure 4 (top-figure) then indicates that the initial mass of these progenitors was ≥ 30-40 $M_\odot$. In particular, XN Mon 75 and XN Cyg 89 contain a very massive BH component (10-15 $M_\odot$). Accounting for the mass that was lost during the SN explosion, it follows from Figure 4 (top-figure) that their progenitors had an initial mass in excess of 40 $M_\odot$ (possibly as large as 60-70 $M_\odot$). This means that the evolution of these progenitors has been heavily influenced by stellar wind mass loss during an LBV and RSG phase, similar as the progenitor of the BH in Cyg X-1. Brown et al. (1999) propose a case C type of evolution to explain these systems. However they deny the effects of stellar wind mass loss during the LBV and/or RSG phase of a massive star (Sect. 3 and Sect. 3.1). As a consequence of these SW phases, Case C does not happen in binaries with primary mass ≥ 20 $M_\odot$ and we therefore do not support the idea of Brown and Bethe.

Portegies Zwart et al. (1997) explored models which are based on the classical spiral-in formalism and concluded that it is extremely difficult to explain the observed numbers of LMXBs with BH components.

Using the stellar wind mass loss rate formalisms during the LBV phase and during the WR phase discussed above (Equation 2) and accounting for the final masses of massive stars which depend on these formalisms, Vanbeveren et al. (1998a, b) proposed the following model for the progenitors of LMXBs with a BH component: LMXBs with a massive BH component originate from binaries with a primary mass ≥ 40 $M_\odot$ and with a period which is large enough (of the order of a few dozen days) so that LBV stellar wind mass loss starts before the RLOF (and thus before classical spiral-in) happens (Sect. 3). The effect of RLOF and of the spiral-in is largely reduced.

Here, we propose a second channel: LMXBs with (not too massive) BH components originate from Case $B_c$ binaries with extreme mass ratio, with primary mass between 20(30) $M_\odot$ and 40 $M_\odot$ (Sect. 3.2), with a period which is large enough that the RSG stellar wind phase starts before the RLOF-spiral-in phase reducing the importance of the latter (Sect. 3.1).

Preliminary calculations reveal that most of the binaries survive this phase, i.e. merging is avoided. This may explain the (relatively) large number of observed LMXBs with a BH component. The periods of LMXBs range between 0.2 days and a few days, which at first glance contradict a stellar wind scenario. However, an LBV or RSG wind is dense and the outflow velocity may be comparable to the orbital velocity of the low mass component. The orbital angular momentum of the outflowing matter can therefore be significantly larger than the momentum of matter at the moment it leaves the loser (Sect. 3) implying a large reduction of the binary period. Moreover, BH formation may be preceded by a SN explosion. If the latter is asymmetrical, even a binary with a pre-SN period of the order of several days can become a post-SN binary with very small period.

## 5. POPULATION NUMBER SYNTHESIS (PNS) OF MASSIVE STARS

The skeletons of most PNS codes with a realistic fraction of interacting binaries are very similar. Binaries with mass ratio $q \leq 0.2$ (low mass component is a normal core hydrogen burning star) evolve through a spiral-in phase. Many of them merge.

Case $B_c$ and Case C systems pass through a common envelope phase where one has to account for the effect of RSG stellar wind mass loss, Case A and Case $B_r$ with $q \geq 0.2$ experience RLOF (whether RLOF is conservative or not is still a matter of debate; mass loss from the system is parameterised).

Post-supernova binaries with a NS or a BH component evolve through a common envelope phase. In the case of merging, a so called Thorne-Zhytkow object is formed.

Differences in the results of massive star PNS performed by different research teams are due to

- differences in the evolutionary calculations of single stars and of close binaries that are used to perform PNS

- differences in the mass and angular momentum loss formalism of a non-conservative RLOF, common envelope and spiral-in process

- differences in the physics of the SN explosion and how the SN explosion of one component affects the orbital parameters of a binary

- differences in the criteria used to decide when a binary will merge or not..

The PNS code of the Brussels team has been described in Vanbeveren et al. (1998a,b,c). Summarising:

- we use evolutionary computations of single stars and of binary components with moderate amount of convective core overshooting as proposed by Schaller et al. (1992), no rotation, with the stellar wind mass loss rate formalisms during the different phases as described above (OB-phase, LBV phase, RSG phase, WR phase)

- the effect of mass accretion on the evolution of the mass gainer is followed in detail. We have a library of over 1000 evolutionary calculations of massive close binaries where the evolution of the mass gainer is followed up to the end of its CHeB phase

- our evolutionary library of single stars and of close binaries contains calculations for different Z values ($0.002 \leq Z \leq 0.02$)

- when mass is leaving the binary during RLOF, we assume that this happens through the second Lagrangian point (matter forms a ring around the system). The angular momentum loss therefore can be computed from first physical principles and is parameter free

- BHs form out of primaries in Case $B_r$ binaries with initial mass larger than 40 $M_\odot$, single stars and primaries of Case $B_c$ and Case C systems with initial mass larger than ~25 $M_\odot$, mass gainers who became single after the first SN explosion and who's mass after mass transfer is larger than ~25 $M_\odot$

- the observed space velocities of pulsars indicate that the SN explosion is asymmetrical. This asymmetry can be translated in terms of the 'kick' velocity of the compact remnant. We study effects of kicks on PNS in full 3-D. The adopted kick velocity distribution is a $\chi^2$-distribution (with a tail of large velocities) and we investigate the consequences for different values of the average velocity. BH formation may be preceded by a SN explosion (Israelian et al, 1999; Nelemans et al., 1999) but our code does not account for the effect

- **merging conditions:** interacting binaries experience either a spiral-in process, a common envelope process, RLOF and mass transfer. All these processes stop when the mass loser has lost almost all its hydrogen rich layers or when both components merge before. After the removal of most of the hydrogen, a helium burning remnant restores its thermal equilibrium and shrinks rapidly. From our evolutionary calculations, we deduce the following relation for the equilibrium radius $R_e$ of the helium remnant ($R_e$ in $R_\odot$) as function of mass M (in $M_\odot$):

$$R_e = 0.000062 M^3 - 0.0049 M^2 + 0.18 M + 0.17 \qquad (6)$$

When a star does not accrete mass, its radius remains constant. When accretion happens, our evolutionary library gives us the radius after accretion when the star restored its equilibrium. *Merging is avoided when the equilibrium radii of both*

*components are smaller than their Roche radii*. Binaries with q ≤ 0.2 at birth or with a NS or BH component evolve through a spiral-in phase. For these binaries we compare the equilibrium radius $R_e$ of the He remnant with the orbital radius of the low mass star, NS or BH. Merging happens when the latter is smaller than $R_e$. In our PNS code we assume that the mergers of two normal stars evolve in a similar way as single stars with mass = the sum of the masses of both components before the mass loss process. However, in most of the cases, merging happens when the primary is a hydrogen shell burning star, i.e. when the helium core is fixed. If the process of merging is similar to the process of accretion of mass (= the mass of the lower mass component), a new star is formed with a helium core that is smaller than the one of a normal single star with mass = the sum of the masses of both components and the evolution of such stars may be different from the evolution of normal single stars.

## 5.1. PNS RESULTS FOR O-TYPE STARS AND WR STARS

In this paper we consider regions where star formation is continuous in time. Starbursts will be considered by Van Bever and Vanbeveren later on. To account for the properties of the observed population of O and early B type binaries, and to account for statistical biases, one is forced to conclude that at least 60% of all massive stars are born as the primary of an interacting binary with orbital period between ~1 day and 10 years (see also Mason and Van Rensbergen, these proceedings). A significant fraction is expected to have mass ratio q ≤ 0.2 where the evolution will be governed by the spiral-in and merger process.

We predict that 15-25% of the O-type stars are post-SN mass gainers of massive close binaries and it was shown by De Donder et al. (1997) that ~40% of the latter class are runaway candidates (i.e. 5-10% of the O-type stars are expected to be runaways due to a previous SN explosion in a binary). Compared to the observed number of runaways, we conclude that at least 50% of them are formed by the SN mechanism in binaries. If the average SN kick velocity = 500 km/s (resp. 150 km/s), less than 20% (resp. more than 50%) of these runaways have a compact companion (NS or BH). These percentages should be very similar in the Galaxy and in the Magellanic Clouds.

The observed overall WR population of the Solar neighbourhood is well reproduced if we start with an initial massive interacting close binary frequency ≥ 60%. We predict a WR+OB binary frequency ~40% and a WN/WC number ratio around 1 corresponding to the observed values. It is interesting to remark that about 40-50% of our predicted WR+OB sample did not experience a classical RLOF with significant mass transfer. They became WR+OB through the LBV and/or RSG stellar wind scenario discussed above. At least 50% of the single WR stars have a binary past (originate from merged binaries or from binary mass gainers where the SN explosion of the primary disrupted the binary) and only few WR stars are expected to have a compact companion (cc) (most of these WR+cc are WR+BHs with a period of the order of days to decades). This explains why very few WR-type (hard) X-ray binaries are observed.

We repeated these computations for low Z regions (Z = 0.002, the metal value of the SMC) assuming a similar OB-type binary frequency as in the Galaxy and the RSG stellar wind mass loss depends on Z. We first used evolutionary calculations where the WR stellar wind mass loss rates are the same as in the Galaxy. PNS predicts that 80% of all WR stars have an OB-type companion (corresponding to observations in the SMC, see C. Foelmi these proceedings), and about 2/3 have experienced RLOF and mass transfer (i.e. only 1/3 were formed by LBV and/or RSG stellar wind mass loss). However, the predicted WN/WC ratio ~1-1.5 whereas the observed SMC ratio ~9. We conclude that a Z-dependence of the RSG mass loss can not be the only reason for the observed difference between the WN/WC number ratio in the Galaxy and the SMC. We then used evolutionary computations where the WR wind is Z-dependent and we obtain a theoretical WN/WC number ratio ~5-10 which corresponds to the observed value. Although SMC statistics of WR stars is small number statistics, we are tempted to conclude that *the variation of the WN/WC number ratio as function of Z is indirect evidence that the WR stellar wind mass loss rates are Z-dependent as predicted by the radiatively driven wind theory*.

Table 2 compares the observed and the theoretically predicted WR/O number ratio for the Solar neighbourhood (SN), the outer Milky Way, the LMC and the SMC. In all cases we assume an interacting close binary frequency ≥ 60%.

*Table 2*   Comparison between the observed and theoretically predicted WR/O number ratio for the Solar neighbourhood (SN), the outer Milky Way, the LMC and the SMC}

| Z | $\left(\frac{WR}{O}\right)_{observed}$ | $\left(\frac{WR}{O}\right)_{predicted}$ |
|---|---|---|
| 0.02 (SN) | ~0.1 | 0.03-0.05 |
| 0.013 (outer MW) | 0.03 | 0.03 |
| 0.008 (LMC) | 0.04 | 0.03-0.04 |
| 0.002 (SMC) | 0.02 | 0.016-0.03 |

It has been noted frequently in literature that in the Solar neighbourhood the WR/O number ratio ~0.1. However, Hipparcos made clear that all O-type distance calibrations are very poor and due to the severe reddening complications in the Milky Way, the number of O-type stars is highly uncertain. PNS predicts a WR/O number ratio ~0.03-0.05, however this value may increase significantly when one accounts for the formation process of O-type stars (Bernasconi and Maeder, 1996; Norberg and Maeder, 2000). Therefore, a comparison between observations and predictions for the Solar neighbourhood seems to be difficult at present. Remark that the correspondence between theory and observation is much better for the outer Milky Way.

The distance-problem is obviously less severe for the Magellanic Clouds and it is interesting to notice that the observed WR/O number ratio in the LMC agrees with the PNS prediction.

The effect of semi-convection on the evolution of massive stars with low Z deserves some attention. During the hydrogen shell burning phase, massive stars expand and may become RSGs. The expansion time scale depends on the treatment of semi-convection and on the effect of metallicity. When semi-convection is treated as a very slow diffusion process (the Ledoux criterion), the expansion is always very rapid and a massive single star becomes a RSG almost at the onset of the CHeB phase. Furthermore, the RLOF of a primary with initial mass ≤ 40 $M_\odot$ in a Case $B_r$ binary is always very fast (independent from Z) and stops at the beginning of CHeB when the primary has lost most of its hydrogen rich layers (resembling a WR star). However, when semi-convection is treated as a very fast diffusion process (Schwarzschild criterion), the expansion timescale is much longer. The models of Meynet et al. (1994) for small Z show that massive single stars (≤ 40 $M_\odot$) straggle around log $T_{eff}$ = 4.2-4 and become RSG only at the end of CHeB. As a consequence RSG stellar wind mass loss plays a very restricted role and single WR stars rarely form out of this mass range. Moreover, since the timescale of the RLOF in Case $B_r$ binaries depends on the expansion timescale during hydrogen shell burning, it follows that in low Z-regions the RLOF of a primary with initial mass ≤ 40 $M_\odot$ in a Case $B_r$ binary lasts for a considerable fraction of the CHeB timescale. The remaining post-RLOF CHeB lifetime is small and this means that only few WR+OB binaries are expected from this mass range. The observed WR/O number value for the SMC ~0.02 whereas PNS predicts 0.016 (semi-convection is fast) or 0.03 (semi-convection is slow). We conclude that the correspondence is satisfactory for the SMC as well.

## 5.2. PNS RESULTS FOR SUPERNOVA RATES

Stellar evolution predicts whether or not a star ends its life with hydrogen rich outer layers. If one knows which stars explode as a supernova, it is therefore possible to use PNS codes to determine the rates of SN type I (no hydrogen) and of SN type II (with hydrogen). This has been done by De Donder and Vanbeveren (1998) with the code described above. One distinguishes type Ia from the other type I (=type Ibc). A type Ia SN may be the result of a White Dwarf in a binary that accretes mass from a main sequence star or red giant that fills its Roche lobe (Ken Nomoto, present proceedings). Type Ibc supernovae progenitors are massive stars that lost their hydrogen by stellar wind mass loss or by RLOF (hypernovae may be a subclass of these type I's) whereas type II progenitors are massive single stars where stellar wind mass loss was not large enough to remove all hydrogen rich layers. It is clear that the relative rates of the different types depend on the binary population (binary frequency, mass ratio and period distribution). We summarise the following three results that were discussed in De Donder and Vanbeveren.

- To determine the *observed* SN II/SN I SN number ratio of one particular galaxy, it is customary to collect data of a large sample of galaxies of the same type. The ratio holding for the whole data set is then taken as representative for all galaxies of this type (Cappellaro et al., 1993a, b; see also Cappellaro present proceedings). However, since the ratio depends critically on the population of binaries, it is clear that this is meaningful only if the binary population is similar in all the galaxies of the data set. Whether this is true or not is a matter of faith.

- Early type (respectively late type) spiral galaxies have an average SN II/Ibc ratio ~ 2.3 (respectively ~5.5). If this difference is due to a difference in the binary population, we conclude that the massive close binary formation rate in late type spirals may be a factor 2 smaller than in early type spirals.

- Consider the data of all galaxies with any type. By comparing the SN II/Ibc number ratio with PNS prediction, we conclude that the average cosmological massive interacting close binary formation rate ~50%.

### 5.3. PNS RESULTS FOR DOUBLE COMPACT STAR BINARIES

PNS predicts the post-massive star collapse population. To estimate the double compact star binary population, let us consider the OB+cc (cc=NS or BH) class binaries. Their further evolution is governed by the spiral-in process of the cc in the OB star and by the various stellar wind mass loss phases of the OB star discussed in Sect. 2. When the binary survives the (spiral-in)+(stellar wind mass loss) phase, the OB star will be a hydrogen deficient CHeB star, resembling a WR star. The further evolution will be governed by WR stellar wind mass loss (Sect. 2) and to predict the population of double compact star binaries it is essential to use CHeB evolutionary computations with the most up to date WR stellar wind rates. Remark that beside the fact that these WR wind rates critically determine whether a massive star ends its life as a NS or as a BH, they also determine the binary period evolution.

The *spiral-in survival condition* has been discussed at the beginning of Sect. 5. In Brussels, we use this survival condition in our PNS code since 1997.

**Results.** PNS computations with the Brussels code have been published by De Donder and Vanbeveren (1998). Summarizing:

- PNS predicts that the birth rate of Galactic double NS systems is about $10^{-6}$-$10^{-5}$/yr, corresponding with observations (van den Heuvel, 1994). Remark however that the observed value is still very uncertain

- mixed systems consisting of a BH and a NS are (2-5) times more frequent than NS+NS binaries

- the formation rate of double BH systems is at least a factor 100 higher than the formation rate of double NS systems

- most of the double NS systems have a period ≤ 2 days which means that most of them will merge within the Hubble time. All double BH binaries and at least 50% of the mixed systems have a period ≥ 10 days. It is questionable whether these systems will merge within the Hubble time.

# 6. ROTATION

The influence of rotation on massive single star evolution has been studied in detail by Meynet and Maeder (2000) (see also A. Maeder, present proceedings). They showed that in order to understand the evolutionary history and the chemistry of the outer layers of individual stars, the effects of rotation are very important.

For PNS research and to investigate the chemical evolution of galaxies, an important question is obviously in how far rotation affects average evolutionary properties.

From the study of Meynet and Maeder, we restrain the following important evolutionary effects.

**The extend of the convective core.** The larger the rotation the larger the convective core during the core hydrogen burning phase and this means that convective core overshooting mimics the effect of rotation on stellar cores. The average equatorial velocity of OB-type stars ~100-150 km/sec (Penny, 1996; Vanbeveren et al., 1998a, b). When this property is linked to the calculations of Meynet and Maeder, we conclude *that the effect of rotation on the extend of the convective core during the core hydrogen phase of massive single stars is on the average similar to the effect of mild convective core overshooting as described in Schaller et al. (1992)*

Due to RLOF a binary component loses most of its hydrogen rich layers and the remnant corresponds more or less to the helium core left behind after core hydrogen burning. Rotation makes larger helium cores so that also here convective core overshooting mimics rotation.

**Mixing of the layers outside the convective core.** Rotation induces mild mixing in all mass layers outside the convective core (rotational diffusion or meridional circulation).

It is a general property that as nuclear burning proceeds, the mass extend of the convective core during core hydrogen burning decreases. In this way a composition gradient is left behind in the stellar interior. Especially in massive stars a situation occurs where these chemically inhomogeneous layers are dynamically stable but vibrationally unstable and this may initiate mixing, a process commonly known as semi-convection (Kato, 1966). Whether this process is fast or slow is still a matter of debate (Langer and Maeder, 1995) but semi-convection and rotational diffusion imply similar effects. Remark however that semi-convection operates only in the inhomogeneous layers whereas rotational diffusion also happens in the outer homogeneous layers. The latter is then responsible for the appearance of CNO elements in the stellar atmosphere but is has little overall evolutionary consequences.

Semi-convection and rotational diffusion decide about the occurence of blue loops in post-core-hydrogen-burning phases, i.e. the blue and red supergiant timescales during core helium burning.

**Rotation and stellar wind mass loss.** The evolution of massive stars is significantly affected by stellar wind mass loss. Question: how does rotation affects the $\dot{M}$? When a theoretical hydrodynamic atmosphere model is used to investigate the effect of $\dot{M}$ on stellar evolution, it is clear that stellar rotation should be included although, at least for stars which are not too close to the Eddington-limit, the effect is at most of the order of 10-20% (Petrenz and Puls, 2000). We like to emphasise that *a study of the effect of rotation on $\dot{M}$ (and thus on stellar evolution) is meaningful only if $\dot{M}$ is determined using a theoretical model*.

However, most massive star evolutionary studies use (semi)-empirical SW rate formalisms like the ones discussed in Sect. 2. Empirical rates are determined by combining observations and atmosphere models. It is clear that these atmosphere models must account for all physical processes, thus also rotation. In this context, it is interesting to notice that empirical SW rates resulting from the analysis of the $H_\alpha$ spectral feature, tend to be too large when 1-dimensional atmosphere models are used (i.e. when rotation is not taken into account) although, again the effect is small for stars that do not rotate close to the break up velocity (Petrenz and Puls, 1996).

**Overall conclusion**

*As far as overall massive single star evolution is concerned, old-fashion semi-convection + mild convective core overshooting mimic the effect of new-fashion rotation.*

**Rotation and PNS.** To study the population of massive stars, present day PNS uses evolutionary calculations with a moderate amount of convective core overshooting (the one proposed by Schaller et al., 1992). I made some trial PNS computations for the Galaxy where evolutionary models are used with varying overshooting (simulating the situation where stars have different rotational properties) and it can be concluded that *rotation hardly changes the conclusions resulting from massive star PNS for the Galaxy where evolutionary calculations are used with moderate amount of overshooting*. The population of blue and red supergiants may be critically affected by rotation but due to the uncertainty in the treatment of semi-convection, it is difficult to draw any firm conclusion.